# A Critical Evaluation of Failure in a Nearshore Outsourcing Project

## What dilemma analysis can tell us

Tony Clear, Bilal Raza and Stephen G. MacDonell
*School of Computing & Mathematical Sciences*
*Auckland University of Technology*
*Auckland, New Zealand*
tony.clear@aut.ac.nz, braza@aut.ac.nz, smacdone@aut.ac.nz

## Abstract

*Global Software Engineering (GSE) research contains few examples consciously applying what Glass and colleagues have termed an 'evaluative-critical' approach. In this study we apply dilemma analysis to conduct a critical review of a major (and ongoing) nearshore Business Process Outsourcing project in New Zealand. The project has become so troubled that a Government Minister has recently been assigned responsibility for troubleshooting it. The 'Novopay' project concerns the implementation of a nationwide payroll system responsible for the payment of some 110,000 teachers and education sector staff. An Australian company won the contract for customizing and implementing the Novopay system, taking over from an existing New Zealand service provider. We demonstrate how a modified form of dilemma analysis can be a powerful technique for highlighting risks and stakeholder impacts from empirical data, and that adopting an evaluative-critical approach to such projects can usefully highlight tensions and barriers to satisfactory project outcomes.*

## 1. INTRODUCTION

In a discussion of the assumed benefits of Global Software Development (GSD), such as *"reduced development costs due to the salary savings possible…access to a larger and better-skilled developer pool…closer proximity to markets and customers"* [1], Conchuir and colleagues sound a note of caution that "anyone engaging in GSD should be aware of the many risks associated with these *"benefits"* [1]. So what can we learn as researchers from a project where those risks have been realized, and their subsequent impacts for stakeholders?

In this paper we conduct a critical review of a troubled Business Process Outsourcing (BPO) project, in which the transition from an onshore to a nearshore service provider has proven to be disastrous. The project concerns the implementation of a system to service the nationwide payroll needs of some 110,000 teachers and associated staff in New Zealand schools [2]. At the time of writing this paper, the project had just been cited as the 'Novopay debacle' by New Zealand Computerworld [3], and it had become the subject of a full ministerial review by a Cabinet Minister allocated specific responsibility for the project, at an estimated cost of NZ$500,000 [4]. An estimate of the costs in additional overtime worked by the 2242 impacted schools in response to the problems occasioned since the system's implementation in August 2012, was in excess of NZ$16 million [5].

Before outlining the history and setting the context for this recent project, some background is given on prior iterations of nationwide education payroll projects some fifteen and twenty years earlier, reviewed by Myers [6] and Gill [7], respectively. The parallels are somewhat depressing. Data from one of these prior studies is used to highlight the wide range of stakeholders involved and the issues they faced. This leads to a brief discussion on the roles of stakeholders and risk assessment in project success and failure [6, 8]. We then discuss how one 'evaluative-critical' [9] research approach, namely a modified form of dilemma analysis [10, 11], may be applied to tease out salient issues and enable significant insight into the inherent tensions that generate risks for such projects.

We subsequently proceed to conduct an exploratory dilemma analysis upon publicly available secondary data taking into account the perspectives of selected key stakeholders in the ongoing Novopay project, which profiles three major dilemma clusters. The paper concludes with a discussion on the merits of this technique within an evaluative-critical research approach [9] and the value of the insights that have been gained.

## 2. HISTORICAL BACKGROUND AND PARALLELS

Major transitions in payroll processing are not new to many large organizations, and are often seen as rather routine projects. Yet in the particular case of the New Zealand Education Department the history of such transitions has been rather chequered. Setting the broader historical context to the project reviewed here may serve to explain a degree of risk aversion on the part of the (now) Ministry of Education, who had experienced a major and embarrassing payroll project failure some twenty years earlier. The project was referred to as *"the failed implementation of a centralised payroll system for the New Zealand Education Department"* [6]. Symptoms included thousands of teachers who found they had not been paid correctly, and hundreds who did not get paid at all on 8 February (the first pay day of 1989); *"relief teachers and some part-time teachers had not been paid by mid-April"* [6]. Yet by June 1989 the Education Department's Director of Management Services "was able to announce publicly that the…computerized payroll system was on target to meet its objective of saving the Government millions of dollars" (through reduced interest costs by avoiding the need to pay lump sum holiday payments in advance). Despite this positive perspective, "less than six months later the centralised payroll processing was scrapped by the government" [6].

Such pronouncements as that made by the Director call into question the definition of project success. Myers, for instance, applying a definition drawing upon "dialectical hermeneutics" (an evaluative-critical research approach), notes: *"Information systems success is achieved when an information system is perceived to be successful by the stakeholders and other observers"* [6]. Yet given the mixed views of the parties to this historical project, questions arise of: who is a stakeholder, and what influence does each have on the outcome? Whose concerns are most likely to be taken into account in the implementation of a new system, and at what stage do they become salient? [12] Stakeholders of a computer system have been defined as: *"People who will be affected in a significant way by or have material interests in the nature and running of the new computerised system"* [Willcocks and Mason (1987), cited in [8]].

Such a broad view of stakeholders is important. For instance, Gotterbarn and Rogerson [8] note the significance of stakeholders beyond the narrow view of the project team and the (ubiquitous) customer as the key stakeholders. They observe in relation to their own case study: "It is the failure to consider these 'extra-project' risks and 'extra-project stakeholders' which make this project a failed one" [8], and argue that inattention to these risks and stakeholders contributes to a high failure rate of newly launched products.

Perhaps an explanation of this all too common phenomenon can be found in the model propounded by Mitchell et al. [13], depicted in Figure 1. Positioning stakeholders in a framework encompassing relative power, legitimacy and urgency of stakeholder concerns enables a mapping of stakeholder types to the case above. For instance the Director of Management Services can be viewed as the *"dominant stakeholder"* pushing his business case for financial savings; the teachers as the *"dependent stakeholders"* hoping to be paid correctly (with the relief and part-time teachers simply waiting to be paid) by the system; and the government (who eventually acted) can be viewed as the *"dormant stakeholder"*, responding to the building pressures imposed by the unsatisfactory operation of the centralized system, and then moving to the position of *"definitive stakeholder"* to effect change.

Nonetheless it must be acknowledged that the system had a highly complex range of stakeholder needs to satisfy, which in itself necessarily raised the risk to the project. From a textual analysis alone of the brief case study description in [6], some twenty-six distinct stakeholders or stakeholder groups were identified. These groups comprised: *Educators, Financial Organizations, Government Departments – National and Regional, non salaried educators, Payroll operational staff, Payroll units, Political, The Press, School Principals and Regional Representatives, Senior Management at National and Regional levels, Teachers' Unions, and the Vendor.*

Yet in 1996 history seemingly repeated itself for the Ministry. It outsourced the operation of its central payroll to Datacom "the largest New Zealand owned IT company" [7]. So this was an onshore outsourcing project, occurring within the volatile context of a large scale government imposed change agenda of radical decentralisation. John Gill (Chairman of Datacom) records the events, "The old system was disintegrating by the day, however, and although the new system was still unsteady, it had to be let run. The new payroll blew apart in a humbling night with teachers not being paid at all. Then over a period of three months, many teachers were overpaid and underpaid, some several times" [7]. After considerable effort and cost, things stabilised after six months of disruption. Gill observed "Lessons learned about the rate of change possible in large systems will never be forgotten by those involved", and "This big bang approach is a recipe for failure" [7].

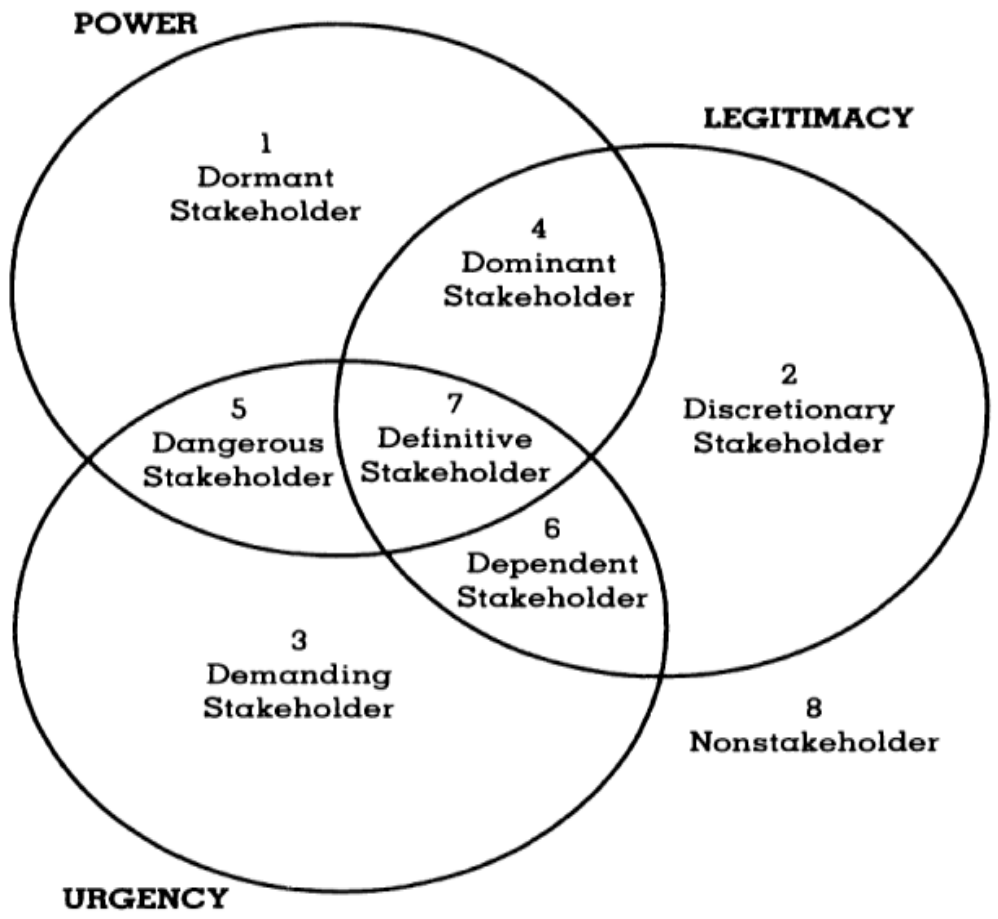

**Figure 1.** Stakeholder typology (ex. Mitchell et al., 1997 [13]).

## 3. CURRENT PROJECT CONTEXT

The project reviewed in this study demonstrates some surprising similarities with the earlier projects respectively of twenty-four and sixteen years ago, perhaps sadly illustrating that (contrary to Gill's prior assertion) we do not learn from our mistakes, or that generational knowledge in implementing software systems does not exceed a ten year timespan?

A chart of the project history by New Zealand Computerworld [3] notes the trajectory of what they term the “Novopay saga” from its initiation in 2005 to its state as at 12 February 2013, wherein “Economic Development Minister Stephen Joyce has now been appointed as Minister responsible for Novopay and has ordered a technical review and a Ministerial inquiry into the troubled payroll system” [3]. Key milestones in the development of the project as outlined by reporter Randall Jackson [3] are highlighted as follows:

- “The Novopay saga began in 2005 when Australian human resources and outsourcing provider Talent2 outbid 10 other potential providers to win a deal to replace the Ministry of Education’s payroll system.
- [The New Zealand Company] Datacom, which was among the unsuccessful tenderers, had provided the service since 1996. It had developed a bespoke system when it took over the ministry’s in-house service”.
- “Talent2 was to provide the payroll service from 2010 with a new service desk, pay clerking service, technology systems and management processes. It would train school staff in the new system and provide on-going support”.
- “*Computerworld* reported in April 2010 that the transition to the new payroll system had been delayed. The new system was due to be operational in the South Island by the middle of the year, and in the North Island toward the end of the year”.
- “[Ministry group manager for the payroll system at the time, Kevin] Wilson said the delays meant that implementation rates were being reviewed… to allow sufficient time for testing to take place”.
- “The original contract, negotiated under the Labour government…, was subsequently renegotiated twice by the National government.
- In 2011 the ministry advertised for a programme director and the go-live date was revised...”
- *“Early in 2011 the go-live date was revised to July 2012 to provide more time to complete customising the payroll system”*, said Fiona McTavish, group manager Education Workforce”.
- “*Computerworld* noted that the project was on the government’s list of high-risk projects”.
- (The project eventually went live in a full nationwide ‘big bang’ implementation on 20 August 2012 [14, p. 16]).
- “*Computerworld* spoke to the Post Primary Teachers Association and the New Secondary Principals’ Council after the first, flawed pay run of Novopay.
- *“We are concerned that the Novopay system has been unable to meet most of its service targets,”* said PPTA president Robin Duff.
- Allan Vester, chairman of the NZ Secondary Principals’ Council, said the issues tended to be around the fringes… *‘situations around leave calculation and relief teachers’”.*
- “*Computerworld* reported on November 15, 2012 that the ministry would end up spending more than [NZ]$100 million on Novopay”. “According to the ministry the total Novopay costs were:
    - Development and implementation - $29.4 million;
    - Long-run cost - $ 12.5 million a year until 2018;
    - Assurity Consulting was paid $350,000 for an initial testing contract. Subsequent contracts took the invoiced total to $842, 000.”

Relevant documents revealed that 147 software defects had been identified when ten project workstream leaders [14, p. 17] and three cabinet ministers signed off Novopay [14, p. 2] and authorized the system to go live in August 2012.

The summary above briefly profiles the history of the troubled transition from one onshore software and payroll services provider to a nearshore BPO provider. A fuller official chronology was made available on the Ministry’s website [15] as part of its response to “Official Information Act requests” (“the data dump”) [3], thereby providing “a comprehensive set of Novopay related information from project inception to current time” [15].

## 4. METHOD

We now consider the project just described, in order to better understand the diversity of stakeholder perspectives and to retrospectively identify sources of risk. In doing so we adopt an ‘evaluative-critical’ research stance. As stated in the classification of research approaches and epistemological positions by Orlikowski & Baroudi [16]:

> *“**Critical** studies aim to critique the status quo, through the exposure of what are believed to be deep-seated, structural contradictions within social systems…criteria we adopted in classifying critical studies were evidence of a critical stance towards taken-for-granted assumptions about organizations and information systems, and a dialectical analysis which attempted to reveal the historical, ideological, and contradictory nature of existing social practices”.*

In the study of Software Engineering research by Glass and colleagues [9], articles which adopted the above definition of [16], i.e. applying an *evaluative-critical* research approach, comprised a mere 1.4% of the 369 research articles analyzed. A more recent survey of the 2009 International Conference on Global Software Engineering proceedings found no instances of papers adopting this approach [17]. Although it is rarely used, we contend that such an approach is ideally suited to an analysis of the current Novopay project, given its major social, political, financial and legal ramifications to what are seen to be standard public sector procurement procedures.

Methodologically we apply an adaptation of dilemma analysis applied in action research studies [10,11], as a useful technique for dialectical analysis. McKernan was sceptical about any ‘researcher imposed’ interpretation based directly on critical social theory. Therefore he proposed a more empirical approach, using the formal theory of contradiction to guide dilemma analysis: *"i.e. that institutions have conflicts of interests, that members are split and divided, and all of this is beset by dilemmas"* [11].

This enables data to be analyzed not in terms of particular *opinions*, but in terms of the *issues* about which opinions were held. The classic procedures for dilemma analysis involve conducting interviews, then analyzing the data in terms of a number of dilemmas, tensions or contradictions (categorized as *ambiguities*, *judgements* and *problems*). A set of perspective documents is then developed, organized by these categories from the perspective of each of the actors in the research project. The dilemmas facing each actor are thus used to build a perspective profile for each role in the research project. These perspective documents are then normally checked with the participants in the research project in order to formulate an overall perspective which transcends individual beliefs. The analysis is thus said to be "a 'mapping' of individual perspectives". The advantage of this method is that it meets one of Melrose's [18] requirements for rigor in action research, namely that interpretations, theories and tentative conclusions be checked with others before, during and after the research process.

However as McKernan [11] observes, developing and cross checking perspective documents is a very time-consuming data analysis method. In this study we use not interviews, but publicly available data, on a high profile and troubled public sector project, as textual sources from which the dilemmas may be elicited. The data resulting from this project may not consist of lengthy interview transcripts, but nonetheless brings forth the concerns of many stakeholders, has surprisingly wide coverage and presents a rich and varied picture. In the interests of efficacy we have elected to apply an abbreviated version of dilemma analysis as a dialectic method to identify the significant tensions that have become apparent in the course of the project. This method may lack the confirmatory strength of a full dilemma analysis, but the dilemmas are grounded in empirical data and commentary on the BPO project, triggered by the insights of the participants at multiple levels (from data entry clerks to Ministers of the Crown), and encompassing the views of the vendor and expert commentators.

The corpus comprised some 65 data sources, many cited here in the references. They range from briefings to ministers; project group meeting records; an official project chronology; press commentaries; and included the Official Information Act 'data dump' response on the Ministry site [15]. Our exploratory analysis has proceeded by focused reading of a representative proportion of these textual sources sentence by sentence, highlighting stakeholders involved and identifying within each sentence of the text any *ambiguity*, *judgement* or *problem* relevant to that stakeholder. Each of these was then tabulated, resulting in 279 instances of an issue identified in this way. The issues were then grouped by stakeholder, stakeholder type, or stakeholder pair where appropriate (e.g. customer/vendor) and refined into issue sets, addressing identified emergent themes. By a process of further abstraction the underlying dilemmas relating to each issue cluster were discerned, named, and the polar dimensions for each dilemma were mapped. After ensuring that each set of stakeholder issues had been mapped to an applicable dilemma cluster, the overall dilemma set was reviewed and mapped against the other stakeholders to produce a consolidated set of clusters and dilemmas which were applicable to more than one stakeholder. This reduced set of dilemmas was then analyzed to clarify both the findings from the study and the efficacy of this somewhat abbreviated method. The extent to which these sources are factual versus mere expressions of opinion may be questioned, but the contradictions that arise from their very diversity have value in more starkly highlighting the embedded dilemmas.

## 5. FINDINGS

A full analysis of the data sources just described would result in a larger set of dilemma clusters than can be represented here. In this section then, we present graphically two primary dilemma clusters relating to the *customer* and *vendor* stakeholder pair to illustrate the findings (as these were the initial groups of stakeholder perspectives and dilemmas supported by the data that emerged as thematic clusters from the analysis). These two dilemma clusters are supplemented by a fuller tabulation (adapted from the approach of Talanquer et al. [19]) of the issues facing *end-user stakeholders* of the system, and the dilemma set relating to *End-user Satisfaction*.

### A. Contract and Service Achievement dilemma cluster

We first present our *Contract and Service Achievement* dilemma cluster. The stakeholders and issues, identified as *problems*, *ambiguities* or *judgements*, from which the cluster was derived, are tabulated in Appendix A. In combination with the dilemmas shown in Figure 2, this illustrates the approach to dilemma analysis and highlights the data-driven nature of the method, despite its critical focus. It also highlights the ability of the method to crystallize higher level tensions applicable to other BPO projects from raw documentary data and commentary on a specific project.

| Left hand pole | Dilemma | Right hand pole |
|---|---|---|
| | **Dilemma Cluster – Contract and Service Achievement** | |
| Do your work and tender fairly on an informed basis | Vendor tendering strategies ⟷ | Conduct 'Back of the envelope calculation' and promise optimistically |
| Award contract to lowest bidder | Design of tendering process ⟷ | 'Qualification Based Selection' |
| Negotiation, definition and measurement of Vendor favorable position | Definition of service to be provided ⟷ | Negotiation, definition and measurement of Customer favorable position |
| Agreed measures in place | Assessment of service achieved ⟷ | No measure in place ('KPI holiday'-service standards relaxed during transition) |
| Targets achieved | Contention over service levels achieved ⟷ | Targets not achieved |
| Planned and agreed remedial action from vendor | Negotiation over service levels achieved ⟷ | No vendor action, Vendor excuses |

**Figure 2.** Contract and Service Achievement dilemma cluster.

As noted, the dilemmas were developed from the data sources and issues extracted in Appendix A, and so are empirically grounded in the data. For instance for the *Vendor tendering strategies* dilemma, Paul Matthews, CEO of the New Zealand Institute of IT Professionals (IITP) has starkly illustrated the dilemma facing IT vendors [20]:

"*So what do you do if you're tendering on a hugely complex project like this one? Well, you basically have two options.*

*You either spend a considerable sum…working through the requirements in absolute detail and properly speccing the whole project out, all the while knowing that there's still a very high chance you'll lose the tender…*

*Or option 2, you basically "wing it". You still spend a bunch of time, but you only end up with a rough idea of how complex you think the project will be, cost that out, build some fat in for complexity you missed, put in a bid and cross your fingers.*"

For the *Design of tendering process* dilemma, IT Professional John Rusk observes in the IITP newsletter [21] that sound practices in engineering design projects apply "Qualification Based Selection" [22] which separates out pricing from capability, in order to support collaboration towards safe, well-designed engineering outcomes with fair rewards. He further cautions: *"After 15 years as an insider, I can say one thing with certainty:* ***awarding contracts to the lowest bidder is optimistic at best, and dangerous at worst".*** [21]

In the case of the *Definition of service to be provided* dilemma, there appeared to be several related issues. The Ministry's stated cost and risk reduction option of outsourcing both the system and the business processing responsibilities to be managed by contract [23] raised its own challenges. In the Minister's 'report back' to Cabinet on the business case for the BPO option, he stated that: "*the contract has defined the quality of service to be provided through a set of key performance indicators*" [23].

Failure on the part of the vendor to meet critical indicators would incur penalties. Yet in the 2013 briefing to the incoming Minister on the, by then, troubled project it was reported that "*the Ministry has no say in how the service is designed – it only specifies the key performance indicators on which the service is judged*" [24].

Thus while the BPO option had been selected to reduce the Ministry's operational risk, the loosely defined contract for service negotiated actually served to increase the risk should the supplier find itself in difficulties. It also gave scope for finger-pointing and argument over the respective responsibilities of the parties.

However there may be times in the life of a project when it is considered more beneficial to ongoing progress for the parties to co-operate. In the case of the *Assessment of service achieved* dilemma the Ministry opted to relax its contractual requirements and grant a "KPI holiday" to the vendor during the messy transition period to avoid a compounding of backlog related issues [25]. This was noted as "very generous given the stress the sector is under because of low service levels" [25]. Yet in respect of the *Contention over service levels achieved* dilemma, the same briefing noted that Talent2 inexplicably did not wish to "agree to the variation" [25]. Coinciding with this offer the ministry proposed "withholding payments to Talent2 because of outstanding software defects" [25]. Two months later the Ministry wrote to Talent2, notifying it that reduced payment penalties would be applied due to failure to meet the contracted critical KPI's for two pay cycles in the previous month [26]. The letter further noted that "Talent2 has provided to the Ministry a draft 'KPI dashboard' on 8 October 2012 but has not otherwise reported against KPI's…the Ministry does not consider that the 'KPI dashboard' meets Talent2's reporting obligations" [26].

In related email correspondence between the Chairman of Talent2 and the Associate Minister for Education, Talent2 replied that: "*…this entire project was a collaborative effort between us and the Ministry and the service and system we have built reflect a specification and model that the Ministry participated in and approved. Some of the design assumptions are clearly being challenged and we are working with the Ministry to readjust. A good example of this would be the many thousands of relievers and the introduction of a new payslip*" [27].

More positively Talent2 noted in the same email the extra resourcing it was dedicating to the project: "over 30 additional FTE have been already added since go-live" [27]; service desk loadings were 20% higher than any day since go-live and call levels would be monitored for trends, with further staff to be added if required; a business analyst who had resigned had been replaced; a plan was in place with the Ministry to begin outbound calling relating to end-of-year processing problems; and senior management planned to visit the following week and then the following month to review the project status. The gist of this communication then, was a mix of constructive responses, and some blame sharing/shifting and excuses.

### B. Global Software Development dilemma cluster

The next customer-vendor dilemma cluster that emerged and is directly relevant to the focus of this paper is the *Global Software Development* cluster, portrayed in Figure 3.

Although GSD was a natural thematic cluster given the focus of the research, these dilemmas are also well-grounded in the data sources, for instance for the *How to better understand customer context* dilemma [28] in which the Talent2 CEO John Rawlinson cries foul about press reports of the $30 million paid to the company, by noting the role of New Zealand partners Fronde and Asparona in the infrastructure provision and software customizations. Contrasting this view is an anonymous blog entry, highlighting the cultural sensitivities between a large neighbour and a small country, in response to Keall's article [28]:

> *"NZ is littered with examples of an Aussie centric organisation of blow hards who relocate / centralise support services across the ditch and find that having to then contract local expertise is far more costly than the local staff they had laid off or other such shining examples of management talent".*

Paul Matthews (IITP CEO) notes how the Government's open tendering *procurement strategy* can disadvantage local firms and hamper development of a local IT industry [20]. The Education Ministry's briefing to the Minister noted that Talent2 "had no *senior management* bandwidth in New Zealand…impeding their ability to make decisions about resourcing, priorities and future improvements to the service" [24]. The business case for the BPO project noted that the Ministry would work closely with the existing service provider to support them in achieving "a profitable end of contract" [23]. But when the Ministry was evaluating contingency options, Leanne Gibson, Chair of the Payroll Reference Group, noted that both the incumbent New Zealand provider and the Australian firm had independently stated "that they would not work with each other as business partners" to provide a shared service [29]. Here we see, even in a nearshore setting, the classic GSD challenges arising from a combination of cultural and geographic distance.

| Left hand pole | Dilemma | Right hand pole |
|---|---|---|
| **Dilemma Cluster-GSD** | | |
| Hire/sub-contract nearshore or offshore | How to better understand customer context ⟷ | Hire/sub-contract 'in-country' |
| Enable onshore vendors (parceling work into manageable projects) | Small country procurement strategy ⟷ | Advantage large offshore vendors (by specifying prime contractor role in v large scale projects) |
| Onshore site | Location of senior vendor management ⟷ | Offshore/nearshore site |
| Period of 'bigamy' (dual vendor with progressive phase-out of incumbent) | Vendor relationship transition ⟷ | 'Serial monogamy' (clean break with incumbent) |
| Service purchaser (lack of control and visibility, technical and contestability risks)<br><br>Service provider(license payroll software, purchase hardware and host at 3rd parties, more expertise and responsibilities | Rationale for new payroll project<br>**Stage 1**<br>⟷<br>**Stage2** | Service provider<br><br>Service purchaser (delegating risk, completely outsourced services, lower cost, higher quality) |

**Figure 3.** GSD dilemma cluster.

Challenges with the *Vendor relationship transition* dilemma, among other factors, resulted in the *clean break* option (or 'big-bang' cutover in this project, slated by IITP CEO Paul Matthews [20]). Yet, in the original Cabinet decision to implement Novopay [30] the then Minister of Education Trevor Mallard had outlined the careful attention that had been given to risk mitigation during transition, with strategies such as: "an eight month staggered implementation that implements the payroll in five distinct areas avoiding a high risk *big bang* implementation" [30]. Somehow over the duration of the project this strategy was abandoned. The BPO contract with Talent2 initially envisaged a two stage cutover "Novopay was to go live in 2010, South Island first, then North Island" [14, p. 9], but eventually the system went live across the whole country at once. The rationale given [14, p. 9] was that "Staggered implementations would add great complexity to the implementation without necessarily significantly mitigating the risks". This argument revolved around the perceived risks associated with continuing with the existing service provider and the fact that many staff worked across schools yet needed to receive a single payslip. In addition, the contingency option negotiated with Datacom, the existing service provider, was due to expire and there was a desire to avoid "the end of year/ start of year window" [14, p.16].

A major dilemma that also emerged in our analysis revolves around the *Rationale for new payroll project*. The legacy system in 2005 was considered high risk, functionally obsolete, was not meeting the new demands of reporting and decision making, was difficult to support and lacking in operational control [14, p. 3, 29], which provided the *Rationale for new payroll project Stage 1.* The New Zealand Cabinet decided to replace this system to enhance business processing and provide online access to end users. One of the stated reasons was to 'regain control' from the external vendor and carry out implementation internally. Yakhlef and Sié [31] describe a common process of organizations moving from producer to purchaser of IT services. In contrast, in Stage 1 of this project the Ministry was required to transition back from purchaser to provider of services. Provision of services requires different skills to those needed in purchasing [31]. This transition approach also coincides with the evolutionary framework provided by Mirani [32] in which he demonstrated through a case study that offshore applications tend to evolve and become business critical and the client may seek to regain control by establishing a "command based hierarchy". Although for the project under consideration the service was not to be offshored – rather, the legacy system was initially to be outsourced onshore – the situation carries the similar implication of having an external provider and the notion of 'losing control'. To regain control and visibility, it was decided that the legacy system should be replaced by a commercially available payroll package and the Ministry would customize it by enhancing the business processing and providing online access to principals and administrators at schools [30]. It was soon realized (in 2007) that this approach might not be optimal or feasible after all [33,34]. After a reassessment, the project's Steering Committee recommended a Business Process Outsourcing (BPO) approach. This approach was considered to be less costly, transferring the risk to an external vendor and requiring less expertise and fewer resources of the Ministry – and provided the *Rationale for new payroll project Stage 2*. Hence, the project, that was originally initiated to regain control, ended up again being outsourced to an external vendor.

Comparing this approach to the earlier mentioned evolutionary framework of Mirani [32], the relationship between vendor and client often starts with contract-based assignments in which elementary level projects are outsourced to external vendors. Over a period of time the client assigns complex applications to selected vendors which cultivates network-based relationships; we argue that the relationship between Datacom and the Ministry during the course of the legacy system operation was network-based due to the complexity and nature of the relationship. When applications become business critical, the client typically regains control by transitioning from network-based to a command-based hierarchy. In *stage 1* of this project that transition was reversed. The Ministry would regain control as a provider by replacing the legacy system with an internally customized commercially available system. Only the readily defined payroll processing services would be contracted. Figure 4 further elaborates this process: in *Stage 1*, we see this reverse approach to the model of Mirani [32] being followed by the Ministry; however, in *Stage 2*, in a plan to reduce cost and transfer risk, the project reverts to the Mirani model through outsourcing to a new external vendor.

It could be a characteristic of outsourced Government projects that contract and network models (perhaps “in conjunction with long-term contracts” [32]) will necessarily predominate. The forms of control through a “command and control hierarchy” open to a firm [32], such as “acquiring a formal stake in a vendor organization through part or full ownership” or setting up “a captive offshore subsidiary” are usually less palatable for Government organizations. If ideological reasons do not mitigate against such arrangements, natural risk aversion or shortage of capital are often barriers.

### C. End-User Satisfaction dilemma cluster

A summary of issues and dilemmas for a third cluster addressing *End-User Satisfaction* has been included to indicate the scope for further analysis (cf. Appendices B & C). One echo of the prior project [6] can be seen in the fate of those stakeholders “not paid” by the system, the *non salaried staff* (Appendix B). As earlier noted in relation to Figure 1 above, this group was again relegated to the “dependent” stakeholder category. Their lack of significance can be seen in the comments made by the Talent2 Chairman when he opined that “the many thousands of relievers” [27] presented a surprise, although they were clearly part of the reality and always required to be handled by the payroll system.

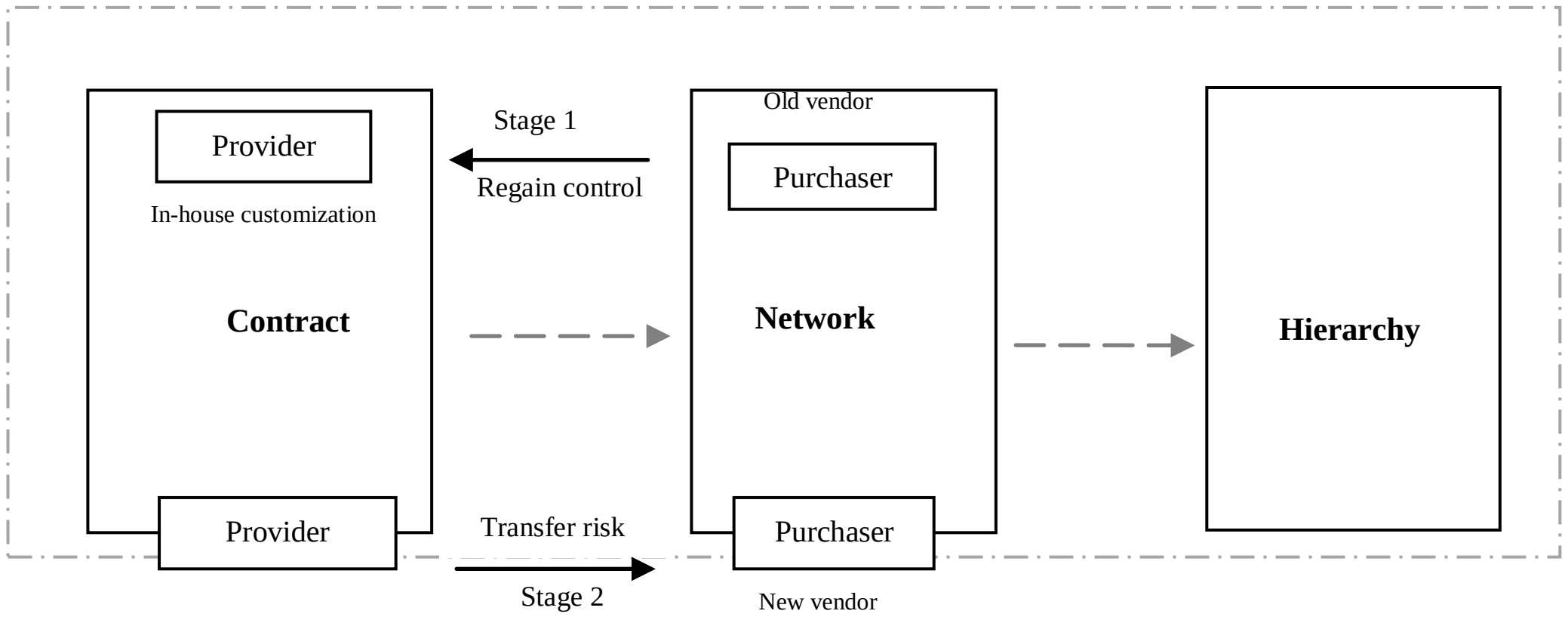

**Figure 4.** Novopay Payroll Transition process.

## 6. LIMITATIONS AND FUTURE WORK

### A. Limitations of Study

The adapted form of dilemma analysis utilized here could be criticized for a lack of focus that might otherwise result from tighter interview based protocols. However, the process has taken a somewhat opportunistic advantage of the plethora of data relating to the project now in the public domain [15]. A future study could adopt a more structured and systematic sampling strategy. Yet the steps of analysis are relatively consistent in progressively highlighting issues relevant to identified stakeholders, framing them as dilemmas and consolidating them with explanatory narratives. The closeness to the situation inherent in an action research project has not been possible in this research. But perhaps that aids the research by ensuring a degree of distance from the context, that a more ‘embedded’ researcher would lack? Triangulation of dilemmas across multiple data sources has not been possible, in the absence of interview data. The fact that the project is a mess, to some extent leaves the data as a mess, but with selective probing it is open to useful analysis. This exploratory study has focused on two stakeholder pairs and one stakeholder group in relative depth, and a wider set of stakeholders and issues have been summarized (in Appendices A & B) to indicate the scope for further analysis. So this study remains in a sense a work in progress, both in scope and method. At this stage the method can be described as largely processual, summarized in diagrammatic form and elaborated in explanatory narratives. Incorporating more robust verification techniques, such as inter-rater reliability assessments, could be a useful future check for consistency of classifications. Nonetheless, some concrete findings have been reported and project-specific conclusions can be drawn which may offer more general insight to others engaged in such projects.

### B. Future Work

As noted above there is scope for a more comprehensive follow up study. A fuller analysis could include an additional interview phase with the key stakeholders to confirm the consistency of stakeholder views upon the dilemma clusters this study had identified. We believe there is also scope for application of the method to a project in progress rather than retrospectively when things have gone awry. Periodic project dilemma audits for instance, could enable focused review and critique of current project issue and risk registers. Given the ready availability of many documentary sources in any software project, there should be data readily available for analysis in this way. Simplifying the method for practitioners to apply effectively will require some work, perhaps aided by a focusing strategy. For instance further research may reveal a set of common dilemmas that may even act as 'archetypes' for consideration.

In some respects the process of dilemma analysis replicates elements of a Software Development Impact Statement or "SoDIS Inspection" [35,36], a qualitative software risk assessment process which adopts a multi-stakeholder perspective. In a SoDIS inspection stakeholders and their concerns are similarly assessed for risks and clustered accordingly. Although not specifically designed with a GSD emphasis the process has scope for accommodating cross border projects (e.g. the nearshore COTS project profiled in [36]). Conducting a SoDIS inspection [35, 36] using the data sources available from this project would be one method by which a follow-up comparison study could be conducted. That could serve to triangulate the findings, and demonstrate greater methodological rigor. In adopting the notion of poles in dilemma analysis the prospect of considering the repertory grid [37] as another possible data collection/analysis method is evident. While these two approaches might not be classified as evaluative-critical in their stance, they could be included in a multi-methodological study. Nonetheless there is a wide range of other analysis techniques that could be adopted within an evaluative-critical approach (cf. for instance [6, 38, 39, and 40]).

## 7. CONCLUSIONS

In this exploratory analysis we have applied an adapted form of dilemma analysis [10] to conduct an 'evaluative-critical' [9] review of a troubled BPO project. The study has focused on selected stakeholders of the project and elicited a set of dilemmas they have encountered. We believe there is value in the technique for GSD and BPO practitioners from the insights we have been able to gain through its application. We also believe that a fuller study addressing all project stakeholders (in excess of the 26 in the earlier project) would enable a more comprehensive analysis of the project and its issues to be presented. One key insight from the sheer number and diversity of issues and dilemmas exposed, is to reinforce the views of Myers that in IS implementation "*there are contradictory perceptions of 'fact', subjective perceptions, and historical factors that shape the context of the implementation effort*" [6]. This strongly supports the use of *evaluative-critical* approaches to research, as they offer methods able to deal with such shifting terrain. We believe that the analysis extends existing GSD research approaches, the findings presented are meaningful and offer empirically grounded insights into GSD and BPO projects that are novel and timely.

## REFERENCES

[1] E. Ó. Conchúir, *et al.*, "Global Software Development: Where are the Benefits?," *Communications of the ACM*, vol. 52, pp. 127-131, 2009.

[2] Ministry of Education. (2013, 18 Feb). Novopay Technical Review – Terms of Reference. Available: http://www.minedu.govt.nz/~/media/MinEdu/Files/TheMinistry/NovopayProject/MinisterialInquiry/TechnicalReviewTermsOfReference.pdf

[3] R. Jackson. (2013, 12 February) Charting the Novopay Debacle. *Computerworld*. Available: http://computerworld.co.nz/news.nsf/news/charting-the-novopay-debacle

[4] H. Rutherford. (2013, 4 February ) Novopay inquiry head named. Stuff.co.nz. Available: http://www.stuff.co.nz/national/politics/8261404/Novopay-inquiry-head-named

[5] C. Mann. (2013, 12 February ) Novopay costs $8m in overtime. Stuff.co.nz. Available: http://www.stuff.co.nz/national/education/8289856/Novopay-costs-8m-in-overtime

[6] M. Myers, "Dialectical hermeneutics: a theoretical framework for the implementation of information systems," *Information Systems Journal*, vol. 5, pp. 51-70, 1995.

[7] J. Gill, "Some New Zealand public sector outsourcing experiences," *Journal of Change Management*, vol. 1, pp. 280-291, 2000

[8] D. Gotterbarn and S. Rogerson, "Responsible Risk Analysis For Software Development: Creating The Software Development Impact Statement," *Communications of the AIS*, vol. 15, pp. 730-750, 2005.

[9] R. Glass, *et al.*, "Research in software engineering: an analysis of the literature," *Information and Software Technology*, vol. 44, pp. 491-506, 2002.

[10] R. Winter, ""Dilemma Analysis": A contribution to methodology for action research," Cambridge Journal of Education, vol. 12, pp. 161-174, 1982/09/01 1982.

[11] J. McKernan, Curriculum Action Research. London: Kogan Page, 1991.

[12] L. McLeod, *et al.*, "A Perspective-Based Understanding of Project Success," *Project Management Journal*, vol. 43, pp. 68-86, 2012.

[13] R. Mitchell, et al., "Toward a Theory of Stakeholder Identification and Salience: Defining the Principle of Who and What really Counts," The Academy of Management Review, vol. 22, pp. 853-886, Oct 1997.

[14] D. Cashmore, "Novopay Chronology," Wellington, Ministry of Education, Internal Memorandum, 5 Dec 2012. Available: http://www.minedu.govt.nz/~/media/MinEdu/Files/TheMinistry/NovopayProject/Background/MemoNovopayChronology.pdf

[15] Ministry of Education. (2013, 18 Feb). Novopay Information Release. Available: http://www.minedu.govt.nz/theMinistry/NovopayProject.aspx

[16] W. Orlikowski and J. Baroudi, "Studying Information Technology in Organizations: Research Approaches and Assumptions," *Information Systems Research*, vol. 2, pp. 1 - 28, 1991.

[17] T. Clear and S. G. MacDonell, "Understanding technology use in global virtual teams: Research methodologies and methods," *Information and Software Technology*, vol. 53, pp. 994-1011, September 2011.

[18] M. Melrose, "Maximising the Rigour of Action Research? Why Would You Want To? How Could You?," Field Methods, vol. 13, pp. 160-180, 2001.

[19] V. Talanquer, et al., "Revealing Student Teachers' Thinking through Dilemma Analysis," Journal of Science Teacher Education, vol. 18, pp. 399-421, 2007/06/01 2007.

[20] P. Matthews. (2013, 12 February 2013) Opinion: Early Lessons from Novopay. Computerworld. Available: http://computerworld.co.nz/news.nsf/news/opinion-early-lessons-from-novopay

[21] J. Rusk. (2013, 20 February) Price-based tendering to blame for Novopay? *Institute of IT Professionals: Newsline*. Available: http://www.iitp.org.nz/newsletter/article/416?utm_source=index

[22] B. Ponomariov and G. Kingsley, "Applicability of the Normative Model of Outsourcing in the Public Sector: The Case of a State Transportation Agency," Public Organization Review, vol. 8, pp. 253-272, 2008/09/01 2008.

[23] Secretary of the Cabinet, "New Zealand Cabinet Minute of Decision," Wellington CAB Min (08) 29/2 2008. Available: http://www.minedu.govt.nz/~/media/MinEdu/Files/TheMinistry/NovopayProject/Background/HistoricalCabinetDocuments/CABMin_08_29_2.pdf

[24] Ministry of Education. (2013, 25January). Novopay Briefing to the Incoming Minister. Available: http://www.minedu.govt.nz/~/media/MinEdu/Files/TheMinistry/NovopayProject/BriefingsCorrespondence/BriefingToTheIncomingMinister.pdf

[25] Ministry of Education, "Memo - Briefing for Lesley Longstone for meeting with Talent 2 on 18 September," Ministry of Education, Wellington 17 September 2012. Available: http://www.minedu.govt.nz/~/media/MinEdu/Files/TheMinistry/NovopayProject/BriefingsCorrespondence/BriefingforCEO17Sept2012.pdf

[26] R. Elvey, "Notification of Failure to Meet Critical Key Performance Indicator," Wellington, Correspondence Ministry of Education to Talent 2, 20 November 2012.

[27] A. Banks, "Novopay," Melbourne, Email correspondence Chairman Talent 2 to Associate Minister of Education,15 November 2012. Available: http://www.minedu.govt.nz/~/media/MinEdu/Files/TheMinistry/NovopayProject/BriefingsCorrespondence/EmailAndrewBanksCraigFossNov12.pdf

[28] C. Keall. (2012, 12 November ) Talent2 boss – we're losing money on Novopay. National Business Review. Available: http://www.nbr.co.nz/opinion/talent2-CK

[29] L. Gibson, "Recommendations Associated with the outcome of the Warning Letter to Talent 2 issued on the 5th April ", Wellington, Ministry of Education, Internal Memorandum19 April 2012.

[30] Secretary of the Cabinet, "New Zealand Cabinet Minute of Decision," Wellington CAB Min (05) 20/2 2005. Available: http://www.minedu.govt.nz/~/media/MinEdu/Files/TheMinistry/NovopayProject/Background/HistoricalCabinetDocuments/CABMin_05_20_2.pdf

[31] A. Yakhlef and L. Sié, "From Producer to Purchaser of IT Services: interactional Knowledge," Knowledge and Process Management, vol. 19, pp. 79-90, 2012.

[32] R. Mirani, "Client-vendor relationships in offshore applications development: An evolutionary framework," Information Resources Management Journal (IRMJ), vol. 19, pp. 72-86, 2006.

[33] Secretary of the Cabinet, "New Zealand Cabinet Minute of Decision - Schools Payroll Business Model," Wellington CAB Min (07) 30/3B 2007.

[34] Secretary of the Cabinet, "New Zealand Cabinet Minute of Decision - Replacement of Schools' Payroll Business Case," Wellington CAB Min (07) 43/2 2007.

[35] D. Gotterbarn, *et al.*, "A Practical Mechanism for Ethical Risk Assessment - A SoDIS Inspection " in *The Handbook of Information and Computer Ethics*, K. Himma and H. Tavani, Eds., ed Hoboken, New Jersey: John Wiley & Sons, 2008, pp. 429-472.

[36] C. Kwan, *et al.*, "Refining the SoDIS® Process in the Field: A COTS Project as a Context for Risk Analysis," in *18th Annual NACCQ Conference*. S. Mann and T. Clear, Eds. Tauranga, New Zealand: NACCQ, 2005, pp. 25-32. Available: http://www.citrenz.ac.nz/conferences/2005/papers/choon.pdf

[37] F. B. Tan and M. G. Hunter, "The repertory grid technique: A method for the study of cognition in information systems," MIS Quarterly, pp. 39-57, 2002.

[38] M. D. Myers and H. K. Klein, "A set of principles for conducting critical research in information systems," *MIS Quarterly*, vol. 35, pp. 17-36, 2011.

[39] D. Howcroft and D. Truex, "Special Issue on Analysis of ERP Systems: The Macro Level," The DATABASE for Advances in Information Systems, vol. 32, Fall 2001.

[40] D. Howcroft and D. Truex, "Special Issue on Analysis of ERP Systems: The Micro Level," The DATABASE for Advances in Information Systems, vol. 33, Winter 2002.

# APPENDIX A

**TABLE I.** INDICATIVE LIST OF ISSUES, PROBLEMS AND AMBIGUITIES FOR CONTRACT AND SERVICE ACHIEVEMENT DILEMMA CLUSTER

| **List of issues for Contract and Service Achievement Dilemma Cluster** | **Issue Categories** | | | | | **Stakeholder types** | |
|---|---|---|---|---|---|---|---|
| | D | M | A | N | DC | M | V |
| Ministry had no say about how the services will be designed | X | | | X | | X | |
| Ministry only had to specify the KPI's against which to judge services | X | X | | | | X | |
| Talent2 have inadequate abilities to make management decisions in NZ | | X | | | | | X |
| Ministry has lack of information about the number of errors | | | X | | | X | |
| Talent2 argues that defects were at acceptable level | | | X | | | X | X |
| Talent2 missed two out of four deadlines | | X | | | | | X |
| System had known errors when it was rolled out | | | | | X | X | X |
| Service model and specifications were approved by the Ministry | | | | X | | X | X |
| Talent2 raising concerns about not being able to meet KPI's | | | X | X | | | X |
| Ministry issues a warning that material breach could be issued | | | | | X | X | X |
| Talent2 hints at litigation if material breach is issued | | | | | X | | |
| Multiple vendors are not willing to work in collaboration | | | | X | X | X | X |
| Talent2 only provided draft KPI's | X | X | | | | X | X |
| Talent2 is losing money on this contract | | | X | | | X | X |
| Project was late by two years due to customization | | | X | X | | X | X |
| Talent2 failed to appreciate the complexity of the project | X | | | | | | X |
| NZ Ministry kept on moving the goal posts | | | | X | | X | X |
| Public sector projects are generally awarded to lowest bidders | | | | X | X | X | X |

Issue Categories: Definition=D, Measurement=M, Argumentative=A, Negotiation=N, Decision=DC
Stakeholders: Ministry=M, Vendor/Talent2=V

## APPENDIX B

**TABLE II.** INDICATIVE LIST OF ISSUES, PROBLEMS AND AMBIGUITIES FOR END USER SATISFACTION DILEMMA CLUSTER

| List of issues for end user stake holders | Issue Categories | | | | | Stakeholder types | | | |
|---|---|---|---|---|---|---|---|---|---|
| | F | S | T | Tch | Sv | E | P | S | NS |
| Under paid, over paid or errors or errors in tax deductions | X | | | X | | X | | | |
| Not paid | X | | | X | | | | | X |
| Loss of confidence in the system | | X | | | | | X | X | X |
| The systems lacks in user interaction | | | X | X | | | X | X | |
| Not able to perform QA checks, due to issues in consolidated reports | X | | X | | | | | X | |
| Transition to new system has been difficult, frustrating and stressful | | X | | | | | X | X | |
| Inadequate online training and lack of motivation | | X | | | | | X | X | |
| Lack of support from customer services and their lack of technical knowledge | | | | | X | | X | X | |
| Long awaiting time by CS and call-backs not carried out | | | X | | X | | X | X | |
| CS staff unable to resolve issues | | | | X | X | | X | X | |
| CS staff provide contradictory information | | | | | X | | X | X | |
| Tax code is randomly updated | X | | | X | | X | | | |
| Empty transactions reports and/or data duplication | X | | | X | | | | X | |
| Internal funds are being used to pay their staff | X | | | | | | | X | |
| Over expenditures through extra hours being paid to pay roll admins | X | | X | | | | X | X | |
| Missing family time and holidays | | X | | | | | X | | |
| Issues closed without further progress | | | | X | X | | X | X | |
| Workarounds placed instead of permanent solutions | | | | X | | | X | X | |
| Backlog of defects in operation | | | | X | | X | X | X | X |
| Non delivery of commitments | | X | | | X | X | X | X | |
| Unmet requirements | | | | X | | X | X | X | |
| Missing successive pay periods | X | | | | | | | | X |
| High Error rates | | | | X | | X | X | X | X |

Issue Categories: Financial=F, Socio-Emotional=S, Time=T, Technical=Tch, Service=Sv
Stakeholders: Educators=E, Payroll operational staff=P, Principals=S, Non-salaried Staff=NS

## APPENDIX C

| Left hand pole | Dilemma | Right hand pole |
|---|---|---|
| | **Dilemma Cluster-End User Satisfaction** | |
| Rely on personal information (resolve issues themselves | Handling variable service quality | Cross checking with technical experts (Tech. experts are at the nearshore location) |
| Work late (extra hours, miss family time) | Handling delay from service centers | Go home on time |
| Not satisfactory (time consuming, unable to resolve issues, abandoned calls and contradictory information) | Resolve service errors and complaints | Satisfactory(Issues resolved) |
| Not satisfactory (work around solutions and report errors) | Resolve service system errors | Satisfactory(Permanent solutions) |
| Training completed | Motivating sector to compete online training | Partially completed or not completed |
| Staying within Budget(waiting for issues to be resolved) | Managing budget to pay staff and keep them satisfied | Relocate funds or overspending |

**Figure 5.** End User Satisfaction dilemma cluster.